\pdfoutput=1
\documentclass[conference]{IEEEtran}
\IEEEoverridecommandlockouts
\PassOptionsToPackage{table, x11names, dvipsnames, svgnames}{xcolor}
\usepackage{cite}
\usepackage{bbm}
\usepackage{amsmath,amssymb,amsfonts}
\usepackage{algorithmic}
\usepackage{graphicx}
\usepackage{textcomp}
\usepackage{xcolor}
\usepackage[english]{babel}
\usepackage[acronym,nonumberlist]{glossaries}
\usepackage{glossaries-prefix}
\usepackage{pgfplots}
\pgfplotsset{compat=1.17}
\usepgfplotslibrary{fillbetween}
\usepackage{amssymb}
\usepackage{import}
\usepackage{bm}
\usepackage{siunitx}
\usepackage{multirow}
\usepackage[]{yquant}
\usepackage{hyperref}
\hypersetup{%
linktocpage=true, 
colorlinks=false,
pdfborder={0 0 0}, %pdfstartpage=3, pdfstartview=FitV,%
breaklinks=true, pdfpagemode=UseNone, pageanchor=true, pdfpagemode=UseOutlines,%
plainpages=false, bookmarksnumbered, bookmarksopen=true, bookmarksopenlevel=1,%
hypertexnames=true, pdfhighlight=/O,%hyperfootnotes=true,%nesting=true,%frenchlinks,%
pdftitle={An Empirical Comparison of Optimizers for
Quantum Machine Learning with SPSA-based
Gradients},%
pdfauthor={Maniraman Periyasam},%
pdfsubject={QML},%
pdfkeywords={},%
pdfcreator={pdfLaTeX},%
pdfproducer={LaTeX with hyperref}%
}

\usepackage[inline]{enumitem}

\usepackage[capitalise]{cleveref}

\newacronym[shortplural=GMMs]{GMM}{GMM}{Gaussian mixture model}
\newacronym[shortplural=HMMs]{HMM}{HMM}{hidden Markov model}
\newacronym[shortplural=DNNs]{DNN}{DNN}{deep neural network}
\newacronym[shortplural=SVDs]{SVD}{SVD}{singular value decomposition}
\newacronym[
    prefixfirst={a\ },% prefix used on first use
    prefix={an\ }% prefix used on subsequent use
]{MCTS}{MCTS}{Monte Carlo tree search}
\newacronym[prefixfirst={a\ },prefix={an\ }]{MDP}{MDP}{Markov decision process}
\newacronym{CMDP}{CMDP}{constrained Markov decision process}
\newacronym{RL}{RL}{reinforcement learning}
\newacronym[shortplural=DTs]{DT}{DT}{decision tree}
\newacronym{SMT}{SMT}{satisfiability modulo theories}
\newacronym{IL}{IL}{Imitation Learning}
\newacronym[shortplural=CNNs]{CNN}{CNN}{convolutional neural network}
\newacronym[shortplural=DQNs]{DQN}{DQN}{deep Q-network}
\newacronym{AI}{AI}{artificial intelligence}
\newacronym{PPO}{PPO}{proximal policy optimization}
\newacronym{ML}{ML}{machine learning}
\newacronym{QML}{QML}{quantum machine learning}
\newacronym{NISQ}{NISQ}{noisy intermediate scale quantum}
\newacronym{QC}{QC}{quantum circuit}
\newacronym{VQC}{VQC}{variational quantum circuit}
\newacronym{VQA}{VQA}{variational quantum algorithm}
\newacronym{MNIST}{MNIST}{modified national institute of standards and technology}
\newacronym{FIM}{FIM}{Fisher information matrix}
\newacronym{IDU}{IDU}{incremental data-uploading}
\newacronym{DRU}{DRU}{data re-uploading}
\newacronym{QRL}{QRL}{quantum reinforcement learning}
\newacronym{SPSA}{SPSA}{simultaneous perturbation stochastic approximation}
\newacronym{SGD}{SGD}{stochastic gradient descent}
\newacronym{QEM}{QEM}{quantum error mitigation}

\begin{document}

%\onecolumn
%\input{reviewReply}

%\setcounter{figure}{0}
%\setcounter{table}{0}
%\twocolumn

\title{An Empirical Comparison of Optimizers for Quantum Machine Learning with SPSA-based Gradients
%{\footnotesize \textsuperscript{*}Note: Sub-titles are not captured in Xplore and should not be used}
\thanks{

* These authors contributed equally (name order randomised). \\
The research is supported by the Bavarian Ministry of Economic Affairs, Regional Development and Energy with funds from the Hightech Agenda Bayern via the project BayQS.\\
email address for correspondence: 
maniraman.periyasamy@iis.fraunhofer.de}
}

\author{
\IEEEauthorblockN{Marco Wiedmann*, Marc Hölle*, Maniraman Periyasamy*, Nico Meyer, 
Christian Ufrecht, Daniel D.\ Scherer, \\ Axel Plinge, and Christopher Mutschler}
\IEEEauthorblockA{\textit{Fraunhofer IIS, Fraunhofer Institute for Integrated Circuits IIS},
Nuremberg, Germany \\\vspace{1mm}}
}

\maketitle

\begin{abstract}
\Glspl{VQA} have attracted a lot of attention from the quantum computing community for the last few years. Their hybrid quantum-classical nature with relatively shallow quantum circuits makes them a promising platform for demonstrating the capabilities of \gls{NISQ} devices. Although the classical machine learning community focuses on gradient-based parameter optimization, finding near-exact gradients for \glspl{VQC} with the parameter-shift rule introduces a large sampling overhead. Therefore, gradient-free optimizers have gained popularity in quantum machine learning circles. Among the most promising candidates is the \gls{SPSA} algorithm, due to its low computational cost and inherent noise resilience. We introduce a novel approach that uses the approximated gradient from \gls{SPSA} in combination with state-of-the-art gradient-based classical optimizers. We demonstrate numerically that this outperforms both standard \gls{SPSA} and the parameter-shift rule in terms of convergence rate and absolute error in simple regression tasks. The improvement of our novel approach over \gls{SPSA} with \gls{SGD} is even amplified when shot- and hardware-noise are taken into account. We also demonstrate that error mitigation does not significantly affect our results.
\end{abstract}


\glsresetall
%%%%%%%%%%%%%%%%%%%%%%%%%%%%%%%%%%%%%%%%%%%%%
%%%%% Sections %%%%%%%%%%%%%%%%%%%%%%%%%%%%%%
\section{Introduction}
\label{sec:introduction}

\IEEEPARstart{M}{achine} learning (ML) has been the key driver for a great variety of recent technological advancements, for example in computer vision \cite{Voulodimos2018}, natural language processing \cite{Otter2017}, automatic translation \cite{Vaswani2017}, self-driving cars \cite{Baheri2020}, and medical diagnostics \cite{fatima2017survey}. However, machine learning algorithms often require a considerable amount of computational resources and data in the training stage. The rapidly emerging field of \gls{QML} aims to tackle these challenges by employing a completely different computational paradigm, namely quantum computing. 

Quantum computing gained a lot of attention since it offers speedup in terms of computational complexity for a range of problems, including factoring numbers \cite{Shor94}, unstructured search \cite{Grover96}, or solving systems of linear equations \cite{Harrow09}. Still, the realization of a fault-tolerant quantum computer is yet to be achieved, with current hardware suffering from low qubit counts, coherence times, and high gate- and readout-error rates \cite{preskill2018quantum}. As these limitations restrict the size of the quantum circuits that can be executed, the quantum computing community turned to hybrid quantum-classical algorithms, such as \glspl{VQA} \cite{cerezo2021variational}, in the hope of surpassing classical computing capabilities. They have been applied in the fields of chemistry \cite{OMalley2016}, nuclear physics \cite{Kiss2022}, combinatorial optimization \cite{Farhi2014}, and machine learning \cite{farhi2018classification}, which will be the focus of this work.

Like in classical ML, choosing the correct optimizer for the learning task is a key factor in keeping computational costs as low as possible. In the current literature, there are two popular choices. On the one hand, one can either use the parameter-shift rule \cite{Schuld19} to evaluate gradients on the quantum hardware and use a gradient-based optimizer, like Adam \cite{ADAM}; on the other hand, one can use gradient-free optimization methods that do not need access to exact derivatives, like \gls{SPSA} \cite{Spall92}. The latter only requires estimating two expectation values per update step, regardless of the number of parameters that need to be optimized in the circuit. In contrast, the parameter-shift rule requires \(2p\) estimations, where \(p\) is the number of parameters.

%In this paper we present the first quantitative comparison of the performance of the Adam, RMSProp AMSGrad %and SGD (with momentum) optimizers with both parameter-shift rule inferred and SPSA approximated gradients. 
In this paper, we present the first quantitative comparison of the following optimizers: (1) Adam, (2) RMSProp, (3) AMSGrad, (4) SGD, and (5) SGD with momentum in conjunction with the gradients estimated using parameter-shift or SPSA rules.
%We evaluate the two approaches 
We evaluate each combination of these approaches on multiple regression tasks from the \textit{scikit-learn} library and on real-world data. Numerical experiments were done on the qasm simulator offered by the IBM Quantum services \cite{ibmq2022}. Both ideal and noisy conditions mimicking the ibmq\_ehningen device are employed to study the effect of noise. Additionally, we examine how the use of error mitigation \cite{Cai22} on the noisy simulation impacts the performance of the optimizers.

The paper is structured as follows: Sec. \ref{sec:related_work} provides an overview of related literature. The relevant theoretical background is introduced in Sec. \ref{sec:Theory}. Sec. \ref{sec:vqc} introduces variational quantum circutis. Sec. \ref{sec:gradient_estimation} explains how gradients can be estimated on quantum hardware. In Sec. \ref{sec:Optimizers} a brief overview of the different optimizers that are used in this paper is given. This is followed by a short high-level introduction to quantum error mitigation in Sec. \ref{sec:error_mitigation}. A detailed description of numerical experiments is given in Sec. \ref{sec:Methods} and subsequently the results thereof are presented in Sec. \ref{sec:Results}.

\section{Related Work}
\label{sec:related_work}
SGD with \gls{SPSA}-based gradients (which is just called \gls{SPSA} in the literature) has recently been compared against other gradient-free and gradient-based optimizers in various settings. The available literature suggests, that \gls{SPSA} is able to outperform other state-of-the-art gradient-based and gradient-free optimization methods in terms of final error, in a wide range of variational quantum computing tasks like finding energy spectra and ground state energies \cite{Sa2023, Bonet2021, Oliv2022, Mihalikova2022} or solving linear systems of equations \cite{Pellow-Jarman2021}, both in noiseless and noisy simulations.

However, the picture is not always that clear. Singh et al. \cite{Singh2022} observed, that the \gls{SPSA} optimizer converged to significantly worse results than all other considered methods when evaluating the ground-state energy and dipole moments of different molecules on a noiseless simulator. While it is affected less by noise than most other optimizers, as can be seen from the improved relative performance to the other methods in the noisy simulation, it is still outperformed by Powell's method, even in the noisy setting.

The core idea of \gls{SPSA}, namely approximating gradients by simultaneous perturbations, has also been used to design new optimizers, that are tailored specifically to the training of variational quantum circuits.
For example, a quantum version of the natural gradient optimization scheme can be built from a second-order variant of \gls{SPSA} that approximates the quantum Fischer information matrix \cite{Gacon2021, Gidi2022}.
%For example \cite{Gacon2021, Gidi2022} use a second-order variant of SPSA to approximate the quantum Fischer information matrix to build a quantum version of the natural gradient optimization scheme.
In addition to that, while SPSA does not require the explicit computation of the loss functions gradient, internally it still computes an approximation to it. Therefore it is possible to use the approximated gradient from \gls{SPSA} with well-established gradient-based optimizers. This idea has been mentioned in \cite{Ebadi2022}, but no quantitative comparisons to standard SPSA or the parameter-shift rule are given. While \cite{Hoffmann2022} provides such a quantitative comparison, it only considers the \gls{SGD} optimizer.

\section{Theoretical Background}
\label{sec:Theory}
To give theoretical background, we start by introducing \glspl{VQC} and continue by presenting the parameter-shift rule and SPSA. Furthermore, we give a detailed list of the optimizers we examine in this work and close with a look at \gls{QEM}.

\subsection{Variational Quantum Circuits}
\label{sec:vqc}
A \gls{VQC} can be formally described as a sequence of parameterized gates $U(\bm{x}, \bm{\theta})$, that prepares a quantum state depending on the initial state $| \psi_0 \rangle$, input $\bm{x}$ and variational parameters $\bm{\theta}$. The output is obtained by computing the expectation value of an observable $A$ \cite{Schuld19}
\begin{equation}
\label{eq:expectation_value}
    f(\bm{x}, \bm{\theta}) := \langle \psi_0 | U^\dagger(\bm{x}, \bm{\theta}) A U(\bm{x}, \bm{\theta}) | \psi_0 \rangle .
\end{equation}
In the following, we will drop the dependency on the input $\bm{x}$ to facilitate the notation.
During the training of this circuit, we want to minimize a loss function $L(f(\bm{\theta}), y)$ with respect to some target values $y$ by tuning the circuit parameters $\bm{\theta}$. For example, this can be realized by iterative gradient descent steps:

\begin{equation}
\label{eq:gradient_step}
    \hat{\bm{\theta}}^{k+1} = \hat{\bm{\theta}}^k - \alpha^k \nabla L(\hat{\bm{\theta}}^k), %\overset{\tiny a)}{=} \hat{\bm{\theta}}^k - \alpha^k \nabla f(\hat{\bm{\theta}}^k)
\end{equation}
where $\hat{\bm{\theta}}^k$ denotes the updated parameter vector, $\alpha^k$ the learning rate and $\nabla L(\hat{\bm{\theta}}^k)$ the gradient at iteration $k$.  
%Without loss of generality we set $L(f(\bm{\theta}), y) = f(\bm{\theta})$ in a), where for more general cases the chain rule can be applied.

\subsection{Gradient Computation}
\label{sec:gradient_estimation}
In order to determine this gradient, we need to compute partial derivatives $\partial_{\theta_i} f(\bm{\theta})$ of the expectation in \cref{eq:expectation_value}, which arise due to the chain rule. Since the classical computational cost of computing the unitary $U$ increases exponentially for larger circuits, it is necessary to obtain gradients directly from the quantum hardware.
%This is where the parameter-shift rule is typically employed to yield estimates of these analytical partial derivatives. %from two estimations of $f(\bm{\theta})$ with shifted parameters. %analytical partial derivatives by computing $f(\bm{\theta})$ twice with shifted parameter $\theta_i$.

%Approaches like automatic gradient computation through backpropagation, as is common in Neural Networks, have no immediate analog in quantum computations, since measurements are irreversible.

\subsubsection{Parameter-shift rule}
The parameter-shift rule enables the computation of exact partial derivatives of \cref{eq:expectation_value} from two expectation values per parameter.
Following refs. \cite{Schuld19} and \cite{Crooks19}, in order to compute the partial derivative \(\partial_{\theta_i}f(\bm{\theta})\) with respect to a single parameter \(\theta_i\), we factor the circuit unitary $U(\bm{\theta}) = V\mathcal{G}(\theta_i)W$ into parts, which are constant w.r.t. \(\theta_i\), and a unitary \(\mathcal{G}(\theta_i)\), which actually depends on \(\theta_i\). Furthermore, we assume that $\mathcal{G}(\theta_i) = e^{-ia\theta_i G}$ with the hermitian generator $G$ having only two distinct eigenvalues $e_0$ and $e_1$. The parameter-shift rule then states: %pauli version schreiben und dann sagen, man kann andere gates entsprechend zerlegen. oder \cite{} showed that it can be generalized for a broader class of gates by decompsing the,...
\begin{equation}
\label{eq:param_shift}
    \partial_{\theta_i} f(\bm{\theta}) = r \left[ f(\bm{\theta} + \bm{s}) -  f(\bm{\theta} - \bm{s})\right],
\end{equation}
with $r = \frac{a}{2}(e_1 - e_0)$ and the components of shift $\bm{s}$ being $s_i = \frac{\pi}{4r}$ and $s_j = 0 \; \forall \; j \neq i$.

From this we conclude that $\partial_{\theta_i} f(\bm{\theta})$ can be computed directly through two expectation value estimations with shifted parameters. Based on this, the whole gradient can be computed by estimating $2p$ distinct expectation values, where $p = |\bm{\theta}|$ denotes the number of parameters in the trainable circuit.

On quantum hardware, expectation values are formed by averaging noisy measurement outputs over multiple circuit runs, called shots. The finite sampling statistics adversely impacts the computed gradient and therefore the whole training process. In order to get a reliable estimate, multiple shots are required per expectation, making this approach intractable even for medium-sized circuits.%In the context of optimization this corresponds to obtaining noisy samples of $L(f(\bm{\theta}), y)$ and calculating their sample mean. 
%method described in Sec. \ref{sec:spsa}, which estimates the gradient solely based on noisy samples of the loss function and randomly perturbing the parameter vector.

%If multiple gates depend on $\theta_i$ the derivative is obtained by shifting each gate separately and summing the results.
%The consideration of gates $\mathcal{G}(\theta_i)$ as described above suffices since a broad class of parameterized gates can be decomposed into products of standard gates for which the parameter-shift rule holds \cite{Crooks19}.
%Besides the similarity of equation \eqref{eq:param_shift} with the finite difference method from numerical optimization, the parameter-shift rule gives the analytical derivative of equation \eqref{eq:expectation_value}. 
% Gavin E. Crooks, Mitarai, Schuld
\subsubsection{Stochastic Perturbation Simultaneous Approximation}
\label{sec:spsa}
\gls{SPSA} is an algorithm designed for circumstances, where only such noisy samples of the loss function are available. It is a gradient-free stochastic optimization algorithm that follows the steepest descent direction on average \cite{Spall98}. %It is a gradient-free stochastic optimization algorithm that enables gradient approximation from noisy samples of the loss function, following the steepest descent direction on average \cite{Spall98}.

To achieve this, the gradient $\nabla L(\hat{\bm{\theta}}^k)$ in the parameter update step (cf. \cref{eq:gradient_step}) is replaced by a gradient approximation $\hat{\bm{g}}^k(\hat{\bm{\theta}}^k) \approx \nabla L(\hat{\bm{\theta}}^k)$, which is obtained by:
% we could remove ^-1 since we are drawing from +1, -1
\begin{equation*}
\label{eq:spsa}
    \hat{\bm{g}}^k(\hat{\bm{\theta}}^k) = \frac{L(\hat{\bm{\theta}}^k + c^k\bm{\Delta}^k) - L(\hat{\bm{\theta}}^k - c^k\bm{\Delta}^k)}{2c^k} \begin{bmatrix}
	(\Delta^k_1)^{-1} \\
	(\Delta^k_2)^{-1} \\
	\vdots \\
	(\Delta^k_p)^{-1}\\
	\end{bmatrix},
\end{equation*}
where $c^k$ denotes a small positive scaling factor and $\bm{\Delta}^k$ a random perturbation vector at iteration \(k\).
The entries $\Delta^k_i$ of the perturbation vector are drawn uniformly and independently from the set $\{-1,1\}$.

\iffalse \cite{Spall92} gives conditions on the convergence, depending on the distribution of the perturbation vector, learning rate $\alpha^k$, scaling factor $c^k$ and the statistical relation of perturbation vector and loss function samples. \fi

We observe that all parameters are perturbed at the same time. This means that the gradient approximation can be done by estimating exactly two expectation values, independent of the number of parameters. This can amount to a drastic reduction in training time through computational savings per parameter update step. However, these savings are only effective if they are not canceled out by an overall increase in updates required to reach convergence. Recent work by Kungurtsev et al.\ suggests, that the overall asymptotic iteration complexity of \gls{SPSA} is the same as of \gls{SGD} \cite{Kungurtsev2022}. They also observe similar convergence rates empirically. This means \gls{SPSA} can provide real, practical speedup in \glspl{VQA}.
%Crucially, recent work suggests, that the overall iteration complexity of \gls{SPSA} is the same as of \gls{SGD} \cite{Kungurtsev2022}, meaning \gls{SPSA} offers a practical speedup over gradient descent with the parameter-shift rule.
%Random search direction only following the direction of steepest descent in average.
%Perturbation vector is i.i.d. following a symmetric Bernoulli distribution where the entries are drawn uniformly from the set $\{-1,1\}$

\subsection{Optimizers}
\label{sec:Optimizers}
The way the parameter update is performed based on the estimated or approximated gradient is determined by an optimizer. A very basic optimizer is the gradient descent step from \cref{eq:gradient_step}. More sophisticated schemes do not only use the current gradient, but keep a record of gradients from previous iterations $H^k = \{\nabla L(\hat{\bm{\theta}}^s)\}^k_{s=0}$ \cite{Choi19}.
%In order to determine how the parameter update is performed based on the estimated or approximated gradient, more sophisticated schemes instead of the basic gradient descent step in \eqref{eq:gradient_step} can be used. 
%An alternative to the basic gradient update steps in \eqref{} are sophisticated schemes called optimizers.
%Optimizer are algorithms that determine how the parameter update is performed during minimization of the loss function $L(f(\bm{\theta}), y)$. Usually, this is based on a set of first-order derivatives or estimates thereof and the loss from current and previous iterations, $H^k = \{\hat{\bm{\theta}}^s, \nabla L(\hat{\bm{\theta}}^s), L(\hat{\bm{\theta}}^s)\}^k_{s=0}$ \cite{Choi19}.
%First order methods
Formally, they can be expressed as
\begin{equation}
    \hat{\bm{\theta}}^{k+1} = \mathcal{M}(H^k, \bm{\phi}^k),
\end{equation}
where the update rule $\mathcal{M}$ gives the updated parameter vector $\hat{\bm{\theta}}^{k+1}$ based on the record $H^k$ and optimizer hyperparameters $\bm{\phi}^k$ e.g. learning rate $\alpha^k$.

The optimizers we examine in combination with parameter-shift rule and SPSA are the following.
% How to cite, just overview paper or cite every original paper
\subsubsection{Stochastic Gradient Descent}
\gls{SGD} uses the same update rule given in \cref{eq:gradient_step}. The stochasticity enters when the gradient is computed over a subset of the training dataset (mini-batches), rather than the entire set. %$\{y_i\}^N_{i=0}$.

\subsubsection{SGD with momentum}
%There are different impelementations
To accelerate SGD a running mean estimate $\bm{\mu}^k$ of the gradient is included, which is updated via
\begin{equation*}
    \bm{\mu}^{k+1} = \gamma \bm{\mu}^k + \nabla L(\hat{\bm{\theta}}^k).
\end{equation*}
This mean is typically called velocity and is intended to accumulate persistent descent directions across parameter updates
\begin{equation}
    \hat{\bm{\theta}}^{k+1} = \hat{\bm{\theta}}^k - \alpha^k \bm{\mu}^{k+1},
\end{equation}
where the momentum $\gamma \in [0, 1]$ is a constant hyperparameter \cite{sutskever2013importance}.

\subsubsection{Adam}
Adaptive moment estimation (Adam) has an individual and adaptive learning rate per parameter, which is based on running estimates of first and second moments
\begin{align*}
    \bm{\mu}^{k+1} &= \beta_1 \bm{\mu}^k + (1 -\beta_1) \nabla L(\hat{\bm{\theta}}^k) \\
    \bm{\sigma}^{k+1} &= \beta_2 \bm{\sigma}^k + (1 -\beta_2) \nabla L(\hat{\bm{\theta}}^k) \odot \nabla L(\hat{\bm{\theta}}^k),
\end{align*}
of the gradient \cite{ADAM}. Here, $\bm{\mu}^k$ denotes the mean estimate and $\bm{\sigma}^k$ the uncentered variance based on the element-wise square $\nabla L(\hat{\bm{\theta}}^k) \odot \nabla L(\hat{\bm{\theta}}^k)$.

For a small number of iterations, these estimates are biased due to their initialisation to zero. To compensate this, correction factors are introduced
\begin{align*}
    \hat{\bm{\mu}}^{k+1} &= \frac{\bm{\mu}^{k+1}}{1 -(\beta_1)^{k+1}} \\
    \hat{\bm{\sigma}}^{k+1} &= \frac{\bm{\sigma}^{k+1}}{1 -(\beta_2)^{k+1}},
\end{align*}
where the decay rates $\beta_1, \beta_2 \in [0,1)$ are taken to the $(k+1)$-th power.
The resulting parameter update is then given by normalizing the mean by the standard deviation $\sqrt{\hat{\bm{\sigma}}^{k+1}}$ and subtracting this normalized mean from the current parameters
\begin{equation}
\label{eq:adam}
    \hat{\bm{\theta}}^{k+1} = \hat{\bm{\theta}}^k - \frac{\alpha \hat{\bm{\mu}}^{k+1}}{\sqrt{\hat{\bm{\sigma}}^{k+1}} + \epsilon}.
\end{equation}
Here, the step size $\alpha$ is a positive scaling factor.

Reddi et al. show that convergence is not generally guaranteed \cite{Reddi19}. Nonetheless, Adam is found to be effective in many practical scenarios.

\subsubsection{AMSGrad}
This optimizer corrects convergence problems of Adam by maintaing the maximum of the second moment over previous iterations
\begin{equation*}
    \hat{\bm{\sigma}}^{k+1}_{\mathrm{max}} = \max(\hat{\bm{\sigma}}^{k}_{\mathrm{max}}, \hat{\bm{\sigma}}^{k+1}),
\end{equation*}
and replacing $\hat{\bm{\sigma}}^{k+1}$ in \cref{eq:adam} with this maximum value \cite{Reddi19}
\begin{equation}
    \hat{\bm{\theta}}^{k+1} = \hat{\bm{\theta}}^k - \frac{\alpha \hat{\bm{\mu}}^{k+1}}{\sqrt{\hat{\bm{\sigma}}^{k+1}_{\mathrm{max}}} + \epsilon}.
\end{equation}

\subsubsection{RMSProp}
This modified version of SGD with momentum keeps a running estimate of the uncentered variance (mean square)
\begin{equation*}
    \bm{\sigma}^{k+1} = \rho \bm{\sigma}^k + (1 -\rho) \nabla L(\hat{\bm{\theta}}^k) \odot \nabla L(\hat{\bm{\theta}}^k),
\end{equation*}
and normalizes the gradient by the standard deviation (root mean square) \cite{tieleman2012lecture}
\begin{equation*}
    \bm{\mu}^{k+1} = \gamma \bm{\mu}^k + \frac{\alpha^k}{\sqrt{\bm{\sigma}^{k+1}} + \epsilon} \nabla L(\hat{\bm{\theta}}^k).
\end{equation*}
%\begin{equation}
%    \hat{\bm{\theta}}^{k+1} = \hat{\bm{\theta}}^k - \bm{\mu}^{k+1}.
%\end{equation}

\subsection{Quantum Error Mitigation}
\label{sec:error_mitigation}
A challenge with near-term quantum hardware is its susceptibility to noise. This noise introduces a bias into expectation values, such as in \cref{eq:expectation_value}. For ML models based on a VQC this bias should be eliminated as much as possible, since their accuracy depends directly on the estimated expectation. 

Methods that reduce noise impact are called Quantum Error Mitigation (QEM) \cite{Cai22}. In contrast to Quantum Error Correction (QEC) \cite{lidar2013qec}, which currently suffers from a large gate overhead, QEM performes post-processing that does not aim to reduce the effect of noise per shot, but over the whole ensemble of shots. Thus, QEM typically comes with a tradeoff, where for a reduced estimator bias the variance is increased \cite{Cai22}. This can be compensated by an increased number of samples, causing a sampling overhead compared to the ideal noise-free scenario.
In this paper, we employ zero-noise extrapolation \cite{temme2017error}, which uses data gathered from boosted noise levels to extrapolate expectation values to the zero-noise limit. It has already been successfully implemented in different variational algorithms \cite{kandala2019error, li2017efficient, dumitrescu2018cloud, kim2023scalable} both in the simulator and on real hardware.
\section{Methods}
\label{sec:Methods}

We conducted numerical experiments to study the performance of different optimizers in training a VQC for multiple regression tasks, where the gradients are either inferred from SPSA or the parameter-shift rule. The following datasets were employed in the experiments:
\begin{itemize}
    \item The \textit{sklearn} python library features a built-in function \textit{make\_regression} (MReg), which generates the regression target from a random linear combination of the input features with some added gaussian noise.
    \item Three different four- to five-dimensional artificial datasets first introduced by Friedman \cite{friedman1991multivariate, breiman1996bagging} (F1, F2, F3). They are built from rational and trigonometric functions of the the input features.
    \item A real world dataset that captures the relationship of the four ambient variables temperature, air pressure, relative humidity and exhaust vacuum pressure with the power output of a Combined Cycle Power Plant (CCPP) \cite{Tufekci2014}.
\end{itemize}
For each dataset, 500 data points were used for training, 50 for validation, and 100 for model testing. The training was done for a maximum of 30 epochs, with one validation step between each epoch. At the end of the training, the model that attained the best validation results was tested on the test dataset.

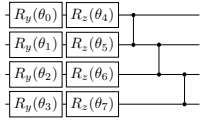
\begin{figure}
    \centering
    \resizebox{0.3\linewidth}{!}{%
    \begin{tikzpicture}
        \begin{yquant}
            qubit {} q[4];
            box {$R_y(\theta_{0})$} q[0];
            box {$R_y(\theta_{1})$} q[1];
            box {$R_y(\theta_{2})$} q[2];
            box {$R_y(\theta_{3})$} q[3];
            box {$R_z(\theta_{4})$} q[0];
            box {$R_z(\theta_{5})$} q[1];
            box {$R_z(\theta_{6})$} q[2];
            box {$R_z(\theta_{7})$} q[3];
            zz (q[1, 0]);
            zz (q[2, 1]);
            zz (q[3, 2]);
        \end{yquant}
    \end{tikzpicture}
    }
    \caption{The variational layer of the VQC. Parametrized \(Y\)- and \(Z\)-rotation gates are used to transform the quantum state. A series of \(ZZ\)-gates with nearest-neighbor connectivity is used to create entanglement between qubits. In the full VQC circuit this variational layer is repeated five times. It is preceded by a single encoding layer.}
    \label{fig:VariationalLayer}
\end{figure}

The function approximator used for these regression tasks is a VQC consisting of four to five qubits, depending on the feature dimension of the considered dataset, with five repeated variational layers (cf. \cref{fig:VariationalLayer}) and a single observable \(A = Z^{\otimes n}\) estimation constituting the output according to equation \eqref{eq:expectation_value}. The classical input data was encoded into quantum states with parametrized single-qubit \(X\)-rotations, where the rotation angles are obtained by linearly scaling the feature space into \(\left[-\pi, \pi\right]^n\). The variational parameters were all initialized to zero. 

All hyperparameters of the optimizers from section \ref{sec:Optimizers} were tuned on each dataset, including the perturbation strength \(c^k\) and learning rate (i.e. a scaling factor applied to each gradient) for the SPSA-based gradient estimation. Since even with SPSA inferred gradients, training the optimizers on the full datasets is computationally expensive, only a subset of 50 randomly sampled points from each dataset were considered for the hyperparameter-tuning.

Since SPSA is often claimed to be especially well suited in noisy settings, it is of prime interest to study the effects of different types of noise, and error mitigation on the optimization procedure. Therefore, the training was done in four different settings. First, an ideal simulation with analytical calculation of all expectation values was performed. Then, shot-noise was taken into account by using a noiseless shot-based simulator with an overall budget of 1024 shots per expectation value estimation. As a next step, realistic hardware noise mimicking the ibmq\_ehningen device was added to the simulation. Finally, the experiments were repeated with zero noise extrapolation \cite{temme2017error} as a technique to mitigate the hardware noise. For this purpose, we used the native error mitigation capabilities of the IBM Quantum services.
\section{Results}
\label{sec:Results}

The results obtained in an ideal, noise-free simulation with the best performing model from the hyperparameter tuning for the SPSA-based gradient estimation step are shown in the top of \cref{tab:Hyperparameter}. Based on the average loss reported in the last row, AMSGrad significantly outperforms SGD, followed closely by Adam and RMSProp when SPSA-based gradients are used. Likewise, the bottom of \cref{tab:Hyperparameter} shows the results obtained during the hyperparameter tuning step of the parameter shift gradient estimation under the same conditions. The average loss shown in \cref{tab:Hyperparameter} already indicates that the parameter-shift based gradient estimation underperforms compared to the SPSA based gradient estimation with all optimizer combinations. However, it should be noted that the hyperparameter results are based on training with 50 data points. Therefore, to validate this observation, we trained all the best performing models again with 500 data points.

\begin{table*}[hbtp!]
    \centering
    \caption{Loss after hyperparameter tuning for each dataset and optimizer in an ideal noise-free simulation using SPSA-based gradients (left) and parameter-shift rule (right).}
    %\begin{tabular}{l|*{5}{c}}
    %    Dataset & SPSA & SGD + Momentum & Adam & AMSGrad & RMSProp\\
    %    \hline
    \resizebox{0.45\linewidth}{!}{
        \begin{tabular}{l|*{5}{c}}
        Dataset & SGD & \shortstack[c]{SGD +\\momentum}{} & Adam & AMSGrad & RMSProp\\
        \hline
        MReg & 0.17 & 0.14 & 0.04 & 0.02	& 0.04\\
        CCPP & 0.13	& 0.14 & 0.11 & 0.10 & 0.10 \\
        F1 & 0.30 & 0.30 & 0.16 & 0.15 & 0.15\\
        F2 & 0.34 & 0.26 & 0.10	& 0.10 & 0.14\\
        F3 & 0.22 & 0.22 & 0.14 & 0.12 & 0.16\\
        \hline
        Average & 0.23 & 0.21 & 0.11 & 0.10	& 0.12 \\
        \end{tabular}
        }
        \vspace{0.5cm}
\resizebox{0.45\linewidth}{!}{
\begin{tabular}{l|*{5}{c}}
        Dataset & SGD & \shortstack[c]{SGD +\\momentum}{} & Adam & AMSGrad & RMSProp\\
        \hline
        MReg & 0.23 & 0.22 & 0.20 & 0.21 & 0.20\\
        CCPP & 0.16 & 0.32 & 0.13 & 0.15 & 0.13 \\
        F1   & 0.30 & 0.30 & 0.22 & 0.24 & 0.23\\
        F2   & 0.41 & 0.45 & 0.38 & 0.39 & 0.39\\
        F3   & 0.33 & 0.36 & 0.26 & 0.27 & 0.26\\
        \hline
        Average & 0.29 & 0.33 & 0.24 & 0.25 & 0.24 \\
        \end{tabular}
    }
    \label{tab:Hyperparameter}
\end{table*}

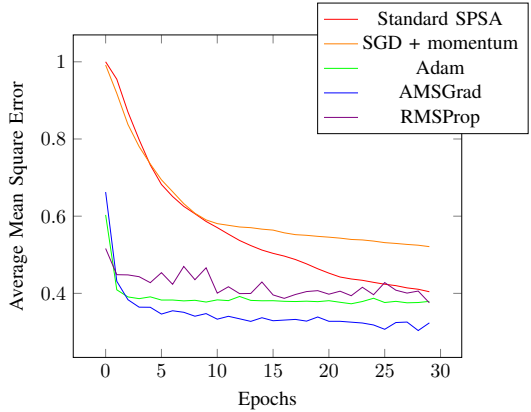
\begin{figure}[htbp!]
    \centering
    \resizebox{0.8\linewidth}{!}{%
    \begin{tikzpicture}
    \begin{axis}[
        xlabel={Epochs},
        ylabel={Average Mean Square Error},
        legend style={at={(0.9, 1.1)},anchor=north},
    ]
        \addplot+[color=red, style=solid, mark=none] table[x=index, y=Standard SPSA, col sep=comma] {img/pgfdata/SPSA_pure_ideal.csv};
        \addplot+[color=orange, style=solid, mark=none] table[x=index, y=SGD + momentum, col sep=comma] {img/pgfdata/SPSA_sgd_ideal.csv};
        \addplot+[color=green, style=solid, mark=none] table[x=index, y=Adam, col sep=comma] {img/pgfdata/SPSA_adam_ideal.csv};
        \addplot+[color=blue, style=solid, mark=none] table[x=index, y=AMSGrad, col sep=comma] {img/pgfdata/SPSA_amsgrad_ideal.csv};
        \addplot+[color=violet, style=solid, mark=none] table[x=index, y=RMSProp, col sep=comma] {img/pgfdata/SPSA_rmsprop_ideal.csv};
        \legend{Standard SPSA, SGD + momentum, Adam, AMSGrad, RMSProp}
        
    \end{axis}
    \end{tikzpicture}
    }
     \caption{Validation results after every epoch during training on an ideal simulator using SPSA-based gradient estimation}
    \label{fig:ideal_validResults}
\end{figure}

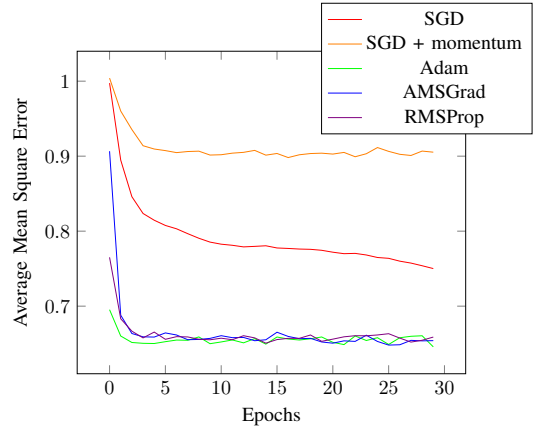
\begin{figure}[htbp!]
    \centering
    \resizebox{0.8\linewidth}{!}{%
    \begin{tikzpicture}
    \begin{axis}[
        xlabel={Epochs},
        ylabel={Average Mean Square Error},
        legend style={at={(0.9, 1.15)},anchor=north}
    ]
        \addplot+[color=red, style=solid, mark=none] table[x=index, y=SGD, col sep=comma] {img/pgfdata/Param-shift_pure_ideal.csv};
        \addplot+[color=orange,  style=solid, mark=none] table[x=index, y=SGD + momentum, col sep=comma] {img/pgfdata/Param-shift_sgd_ideal.csv};
        \addplot+[color=green, style=solid, mark=none] table[x=index, y=Adam, col sep=comma] {img/pgfdata/Param-shift_adam_ideal.csv};
        \addplot+[color=blue,  style=solid, mark=none] table[x=index, y=AMSGrad, col sep=comma] {img/pgfdata/Param-shift_amsgrad_ideal.csv};
        \addplot+[color=violet,  style=solid, mark=none] table[x=index, y=RMSProp, col sep=comma] {img/pgfdata/Param-shift_rmsprop_ideal.csv};
        \legend{SGD, SGD + momentum, Adam, AMSGrad, RMSProp}
        
    \end{axis}
    \end{tikzpicture}
    }
     \caption{Validation results after every epoch during training on an ideal simulator using parameter-shift rule based gradient estimation}
    \label{fig:param_validResults}
\end{figure}

\cref{fig:ideal_validResults} and \cref{fig:param_validResults} show the validation results obtained during training by different optimizers combined with SPSA and parameter-shift based gradient estimation. All curves of the validation results are averaged over five trials and all five datasets. The plots show that the SPSA-based gradient estimation achieves a better solution than the parameter-shift rule in all optimizer combinations. The parameter-shift rule based gradient estimation methods converged to a suboptimal solution and performed two to three times worse than the SPSA-based gradient estimation. On the other hand, the parameter-shift rule requires forty times more circuit simulations per optimization step compared to SPSA, resulting in high computational costs. The same pattern can be seen in the test results in \cref{tab:SPSA_results} and \cref{tab:Parameter_shift_results}. Therefore, no further experiments were performed using the parameter-shift based gradient estimation method. \cref{tab:SPSA_results} shows that AMSGrad achieves a better solution, followed by Adam and RMSProp within the SPSA-based gradient estimation group. As SGD with Momentum in combination with SPSA converged to a suboptimal solution on the validation curve, it was not tested.
%, but SGD with Momentum in combination with SPSA performs worse than standard SPSA. 
From \cref{fig:ideal_validResults} and \cref{tab:SPSA_results}, we can infer that AMSGrad, in combination with SPSA, not only converges two to three times faster but also achieves two times better solutions under ideal simulation.

%\begin{figure}[htbp!]
%    \centering
%    \includegraphics[width=\linewidth]{img/validation_curve_ideal.eps}
%    \caption{Validation results after every epoch during training on an %ideal simulator using SPSA based gradient estimation}
%    \label{fig:ideal_validResults}
%\end{figure}

%\begin{figure}[htbp!]
%    \centering
%    \includegraphics[width=\linewidth]{img/validation_curve_ideal_param-shift.eps}
%    \caption{Validation results after every epoch during training on an ideal simulator using Parameter-%shift rule based gradient estimation}
%    \label{fig:param_validResults}
%\end{figure}

Validation results obtained by different optimizers when simulated on an ideal simulator with shot noise showed similar convergence to ideal simulator results. The test results in \cref{tab:SPSA_results} show that the performance of all optimizers degrades when shot noise is introduced. However, the standard SPSA and SPSA with RMSProp showed a greater decrease in performance compared to SPSA with Adam or AMSGrad. Next, we repeated the same set of experiments with a noisy simulator simulating the ibmq\_ehningen noise model. All the optimizers showed similar convergence patterns during training as shown in \cref{fig:ideal_validResults} and the test results are given in \cref{tab:SPSA_results}. From \cref{tab:SPSA_results}, it can be seen that the performance of standard SPSA seems to degrade under different noise conditions, while SPSA with Adam or AMSGrad exhibited a performance similar to its performance under an ideal simulator with shot noise. Finally, the experiments defined above were repeated with a noisy simulator that simulates the ibmq\_ehningen noise model and the error mitigation method presented in \cref{sec:error_mitigation}. The test results reported in \cref{tab:SPSA_results} show again the same hierarchy, with AMSGrad performing the best on average. However, RMSProp achieves almost the same accuracy on average and even surpasses AMSGrad on the F3 dataset. Additionally, it is striking that with the use of error mitigation all methods perform significantly worse than even in the noisy case. A possible explanation for this is the fact that a global observable \(A = Z^{\otimes n }\) was used to constitute the output of the VQC. Kim et al.~\cite{kim2023scalable} demonstrated that zero-noise extrapolation performs poorly on global observables.

%Average is computed over 5 trials per dataset and the total average is given by the average over all trials of all datasets. Convergence gives the number of epochs till convergence is reached, averaged over 5 trials per dataset and rounded to next integer.
\begin{table}[hbtp!]
    \centering
    \caption{SPSA results. For each individual dataset the reported loss is the average computed over 5 trials. The Normalized average error with respect to standard SPSA (Norm. Avg.), computed over all trials for all datasets, is given at the bottom of each method.}
\resizebox{\linewidth}{!}{
\begin{tabular}{l|l|c|c|c|c}
%\hline
Method & 
Dataset &
SGD &
%\multicolumn{3}{ c| } {SGD + Momentum}  &
Adam &
AMSGrad &
RMSProp \\
\hline
\multirow{6}{*}{Ideal} 
 & MReg & 0.0465 & 0.0317 & 0.0269  & 0.0394\\
 & CCPP & 0.1034 & 0.1028 & 0.1008 & 0.1036\\
 & F1 & 0.1587 & 0.1183 & 0.1436 & 0.1449\\
 & F2 & 0.1327 & 0.1122 & 0.0981 & 0.0841\\
 & F3 & 0.1440 & 0.1441 & 0.1298 & 0.1559\\
 \cline{2-6}
 & Norm. Avg. & 1.0000 & 0.8536 & 0.8198 & 0.8958\\
 \hline
\multirow{6}{*}{\shortstack[l]{Shot-\\based}} 
 & MReg & 0.0492 & 0.0457 & 0,0398 & 0.0546\\
 & CCPP & 0.1107 & 0.1075 & 0,1049 & 0.1014\\
 & F1 & 0.1722 & 0.1292 & 0.1391 & 0.1634\\
 & F2 & 0.1348 & 0.1081 & 0.1146 & 0.1202\\
 & F3 & 0.1551 & 0.1510 & 0.1328 & 0.1646\\
 \cline{2-6}
 & Norm. Avg. & 1.0000 & 0.8849 & 0.8540 & 0.9851\\
 \hline
\multirow{6}{*}{Noisy} 
 & MReg & 0.0773 & 0.0537 & 0,0480 & 0.0622\\
 & CCPP & 0.1219 & 0.1136 & 0.0993 & 0.1070\\
 & F1 & 0.2441 & 0.1719 & 0.1213 & 0.2126\\
 & F2 & 0.1620 & 0.1167 & 0.1005 & 0.1218\\
 & F3 & 0.1762 & 0.1703 & 0.1226 & 0.1655\\ 
 \cline{2-6}
 & Norm. Avg. & 1.0000 & 0.8036 & 0.6498 & 0.8492\\
 \hline
\multirow{6}{*}{\shortstack[l]{Error-\\Mitigated}} 
 & MReg & 0.2080 & 0.1213 & 0.0875 & 0.0885\\
 & CCPP & 0.1589 & 0.1577 & 0.1343 & 0.1388\\
 & F1   & 0.2999 & 0.1780 & 0.1972 & 0.2393\\
 & F2   & 0.4421 & 0.2002 & 0.1800 & 0.1808\\
 & F3   & 0.2804 & 0.2062 & 0.2162 & 0.1941\\
 \cline{2-6}
 & Norm. Avg. & 1.0000 & 0.6716 & 0.6203 & 0.6397\\
 %\hline
\end{tabular}
}
    \label{tab:SPSA_results}
\end{table}

\begin{table}[hbtp!]
    \centering
    \caption{Parameter-shift rule results. For each individual dataset the reported loss is the average computed over 5 trials. The Normalized average error with respect to standard SPSA (Norm. Avg.), computed over all trials for all datasets, is given at the bottom of each method.}
    \resizebox{\linewidth}{!}{\begin{tabular}{l|l|c|c|c|c}
%\hline
Method & Dataset & SGD & Adam & AMSGrad & RMSProp\\ \hline
\multirow{6}{*}{Ideal} 
 & MReg & 0.2096 & 0.2112 & 0,2071 & 0.2107\\
 & CCPP & 0.2208 & 0.1271 & 0.1287 & 0.1269\\
 & F1 & 0.2595 & 0.2352 & 0.2282 & 0.2176\\
 & F2 & 0.3722 & 0.3784 & 0.3754 & 0.3783\\
 & F3 & 0.3017 & 0.2795 & 0.2859 & 0.2840\\
 \cline{2-6}
 & Norm. Avg. & 1.0000 & 0.8866 & 0.8813 & 0.8753 %\hline
\end{tabular}}
    \label{tab:Parameter_shift_results}
\end{table}

\iffalse
\begin{table*}[hbt]
    \centering
    \caption{Parameter-shift rule results, where Average (Avg.) is computed over 5 trials per dataset and Total Average (Tot. avg.) computed over all trials for all datasets.}
    \resizebox{0.8\textwidth}{!}{\begin{tabular}{l|l|*{2}{l}|*{2}{l}|*{2}{l}|*{2}{l} }
%\hline
\multicolumn{2}{ c| } {} &  
\multicolumn{2}{ c| } {SGD} &
\multicolumn{2}{ c| } {Adam}  &
\multicolumn{2}{ c| } {AMSGrad} &
\multicolumn{2}{ c } {RMSProp} \\
\hline
Method & Dataset & Avg. & Tot. avg. & Avg. & Tot. avg. & Avg. & Tot. avg. & Avg. & Tot. avg.\\ \hline
\multirow{5}{*}{Ideal} 
 & MReg & 0.2096 & 0.2727 & 0.2112 & 0.2463 & 0,2071 & 0.2451 & 0.2107 & 0.2435\\
 & CCPP & 0.2208 & & 0.1271 & & 0.1287 & & 0.1269 & \\
 & F1 & 0.2595 & & 0.2352 &  & 0.2282 & & 0.2176 & \\
 & F2 & 0.3722 & & 0.3784 &  & 0.3754 & & 0.3783 & \\
 & F3 & 0.3017 & & 0.2795 &  & 0.2859 & & 0.2840 & \\ %\hline
%\multirow{5}{*}{Shot-based} 
% & MReg & 0.0492& 0.1244 & 25 & 0.0457 & 0.1083 & 8 & 0,0398 & 0,1062 & 8 & & &\\
% & CCPP & 0.1107 & & 29 & 0.1075 & & 5 & 0,1049 & & 11 & & &\\
% & F1 & 0.1722 & & 30 & 0.1292 & & 18 & 0,1391 & & 13 & & &\\
% & F2 & 0.1348 & & 29 & 0.1081 & & 15 & 0,1146 & & 9 & & &\\
% & F3 & 0.1551 & & 29 & 0.1510 & & 8 & 0,1328 & & 13 & & &\\
\end{tabular}}
    \label{tab:Parameter_shift_results}
\end{table*}
\fi
Throughout all methods used, AMSGrad shows the strongest performance. Adam comes close in terms of performance and requires fewer training steps for some datasets such as CCPP. Generally, both optimizers require drastically fewer training steps until convergence is achieved than standard SPSA and SGD with momentum.

\begin{figure*}[htbp!]
    \centering
    \resizebox{0.95\textwidth}{!}{
    \begin{tikzpicture}
        \begin{yquant}
            qubit {} q[4];
            box {$R_x(\theta_{0})$} q[0];
            box {$R_x(\theta_{1})$} q[1];
            box {$R_x(\theta_{2})$} q[2];
            box {$R_x(\theta_{3})$} q[3];
            box {$R_z(\theta_{4})$} q[0];
            box {$R_z(\theta_{5})$} q[1];
            box {$R_z(\theta_{6})$} q[2];
            box {$R_z(\theta_{7})$} q[3];
            
            box {$R_x(\theta_{8})$} q[2]|q[3];
            box {$R_x(\theta_{9})$} q[1]|q[3];
            box {$R_x(\theta_{10})$} q[0]|q[3];

            box {$R_x(\theta_{11})$} q[3]|q[2];
            box {$R_x(\theta_{12})$} q[1]|q[2];
            box {$R_x(\theta_{13})$} q[0]|q[2];

            box {$R_x(\theta_{14})$} q[3]|q[1];
            box {$R_x(\theta_{15})$} q[2]|q[1];
            box {$R_x(\theta_{16})$} q[0]|q[1];

            box {$R_x(\theta_{14})$} q[3]|q[0];
            box {$R_x(\theta_{15})$} q[2]|q[0];
            box {$R_x(\theta_{16})$} q[1]|q[0];
            align -;
            box {$R_x(\theta_{17})$} q[0];
            box {$R_x(\theta_{18})$} q[1];
            box {$R_x(\theta_{19})$} q[2];
            box {$R_x(\theta_{20})$} q[3];
            box {$R_z(\theta_{21})$} q[0];
            box {$R_z(\theta_{22})$} q[1];
            box {$R_z(\theta_{23})$} q[2];
            box {$R_z(\theta_{24})$} q[3];
            
        \end{yquant}
    \end{tikzpicture}
    }
    \caption{The variational layer of the least expressive VQC proposed by Sim et al.}
    \label{fig:MostExpressive}
\end{figure*}

\subsection{Generalization on different architecture ansatz}
To investigate the dependence of the advantage brought in by AMSGrad on the circuit ansatz, we trained two different VQCs with different circuit ansatzes in the same setting with AMSGrad and the standard SPSA optimizer. Since the difficulty of training the circuit depends strongly on the expressivity, the least and most expressive circuits proposed by Sim et al.~\cite{Sim2019, dragan2022quantum} respectively were chosen to validate the performance of AMSGrad in combination with SPSA-based gradient estimation. \cref{fig:leastExpressive} represents one variational layer of the least expressive circuit, and \cref{fig:MostExpressive} represents one variational layer of the most expressive circuit. Five such layers were repeated in each VQC. The results shown in \cref{tab:expressivity_results} validate that the combination of AMSGrad with SPSA-based gradient estimation yields superior results compared to the standard SPSA method, irrespective of the expressivity of the circuit.

%\textcolor{blue}{To investigate the dependence of the advantage brought in by AMSGrad on the circuit ansatz, we trained two different VQCs with different circuit ansatzes in the same setting with AMSGrad and standard SPSA optimizers. The selected circuits were the least expressive circuit and the most expressive circuit proposed by Sim et al.~\cite{Sim2019, dragan2022quantum}. The reason for this selection is to validate the performance of AMSGrad in combination with SPSA-based gradient estimation over different circuits with different expressiveness, leading to different ease of training and gradient descent to a solution. \cref{fig:leastExpressive} represents one variational layer of the least expressive circuit, and \cref{fig:MostExpressive} represents one variational layer of the most expressive circuit. Five such layers were repeated in each VQC. The results shown in \cref{tab:expressivity_results} validate that the combination of AMSGrad with SPSA-based gradient estimation yields superior results compared to the standard SPSA method, irrespective of the expressivity of the circuit.
%}

\begin{figure}
    \centering
    \resizebox{0.3\linewidth}{!}{%
    \begin{tikzpicture}
        \begin{yquant}
            qubit {} q[4];
            H q[0];
            H q[1];
            H q[2];
            H q[3];
            zz (q[3, 2]);
            zz (q[2, 1]);
            zz (q[1, 0]);
            align -;
            box {$R_y(\theta_{0})$} q[0];
            box {$R_y(\theta_{1})$} q[1];
            box {$R_y(\theta_{2})$} q[2];
            box {$R_y(\theta_{3})$} q[3];
            
        \end{yquant}
    \end{tikzpicture}
    }
    \caption{The variational layer of the least expressive VQC proposed by Sim et al.}
    \label{fig:leastExpressive}
\end{figure}

\begin{table}[hbtp!]
    \centering
    \caption{Perormance of AMSGrad and SPSA on circuits with different expressivity}

\begin{tabular}{l|l|c|c}
%\hline
Circuit & 
Dataset &
SGD &
%\multicolumn{3}{ c| } {SGD + Momentum}  
AMSGrad  \\
\hline
\multirow{6}{*}{\shortstack[l]{Least \\ Expressive \\ circuit}} 
 & MReg & 0.2166 & 0.0453 \\
 & CCPP & 0.1406 & 0.1142 \\
 & F1 & 0.2839 & 0.1671 \\
 & F2 & 0.5589 & 0.1287 \\
 & F3 & 0.5147 & 0.1652 \\
 \cline{2-4}
 & Norm. Avg. & 1.0000 & 0.4322 \\
 \hline
\multirow{6}{*}{\shortstack[l]{Most \\ Expressive \\ circuit}} 
 & MReg & 0.2192 & 0.0920 \\
 & CCPP & 0.2353 & 0.1236 \\
 & F1 & 0.3099 & 0.2933 \\
 & F2 & 0.5600 & 0.1533 \\
 & F3 & 0.5483 & 0.5149 \\
 \cline{2-4}
 & Norm. Avg. & 1.000 & 0.6208 \\
\end{tabular}

    \label{tab:expressivity_results}
\end{table}
\section{Conclusion}

This paper proposes an optimization scheme for variational quantum circuits by combining the gradient estimate obtained from simultaneous perturbation stochastic approximation with gradient-based optimizers like SGD, Adam, AMSGrad, or RMSProp. We demonstrate with noiseless simulations on simple regression tasks that using the SPSA-approximated gradient improves both the convergence rate by a factor of three and the final absolute error by a factor of more than two compared to the parameter-shift rule. We further observe that the combined SPSA-AMSGrad optimizer consistently outperforms all other methods, including standard SPSA.

Adam, AMSGrad, and RMSProp clearly outperform SGD for SPSA-inferred gradients when considering shot- and hardware noise. The gap in performance grows to a factor of 1.5 for the full noise model. The performance boost between methods remains the same even when error mitigation addresses the hardware noise.

We conclude that combining the computationally cheap to obtain gradient estimate of SPSA with modern gradient-based optimizers drastically speeds up the training process of VQCs and leads to improved convergence.

% funding is already mentioned on the first page
%The research presented in this paper is supported by the Bavarian Ministry of Economic Affairs, Regional Development and Energy with funds from the Hightech Agenda Bayern.
%%%%%%%%%%%%%%%%%%%%%%%%%%%%%%%%%%%%%%%%%%%%%
%%%%%%%%%%%%%%%%%%%%%%%%%%%%%%%%%%%%%%%%%%%%%
%\let\oldthebibliography=\thebibliography
%\let\endoldthebibliography=\endthebibliography
%\renewenvironment{thebibliography}[1]{%
%\begin{oldthebibliography}{#1}%
%\setlength{\parskip}{1ex plus 1ex minus 0.5ex}%
%\setlength{\itemsep}{6pt}%
%\setstretch{1.0} % ZEILENABSTAND
%}%
%{%
%\end{oldthebibliography}%
%}
\newpage
\bibliographystyle{IEEEtran}
\bibliography{references}

\end{document}